\documentclass[a4paper,11pt]{article}
\usepackage{pos}
\usepackage{lineno}

\title{MID: a muon detector for the ALICE 3 upgrade project }

\ShortTitle{MID: a muon detector for the ALICE 3 upgrade project }

\author*[a]{Jesús Eduardo Muñoz Méndez}

\onbehalf{for the ALICE Collaboration}

\affiliation[a]{Instituto de Ciencias Nucleares, UNAM\\
 Circuito exterior s/n Ciudad Universitaria, 04510 Mexico City, Mexico}

\emailAdd{jesus.munoz@cern.ch}

\abstract{The success of the ALICE programme has allowed to study the hottest state of matter created in the laboratory: the strongly-interacting quark–gluon plasma (sQGP). However, some fundamental questions on the sQGP and other aspects of the strong interaction will remain open even after LHC Run 4. To address this, the ALICE collaboration has proposed a major upgrade of its detector, known as ALICE 3. Among the subdetectors considered for this upgrade is the muon identification detector (MID), whose aim is to contribute to the study of charmonium and exotic-hadron production in heavy-ion collisions. While the baseline design consists of plastic-scintillator-based chambers placed around a cylindrical magnetic iron absorber, different detector technologies are under evaluation for the MID, including multi-wire proportional chambers.

This proceeding presents the results of two MID prototypes based on FNAL-NICADD scintillator bars, Kuraray fibers and Hamamatsu SiPMs, characterized in test beams at the CERN Proton Synchrotron with 3~GeV/$c$ muon- and pion-enriched beams. The muon efficiency as well as the hadron suppression obtained using machine learning techniques will be discussed.}

\FullConference{
43rd International Conference on High Energy Physics (ICHEP 2026)\\
30 July to 5 August, 2026\\
Natal, Brazil
}

\begin{document}
\maketitle

\section{Introduction}

The Muon Identification Detector (MID) is the outermost subsystem proposed for ALICE 3, a completely new apparatus planned for LHC Run 5~\cite{ALICE:2022wwr,Dainese:2925455}, providing access to charmonium and exotic states. The MID will tag muons after a cylindrical hadron absorber, providing acceptance for $J/\psi$ down to zero momentum at midrapidity, which requires a muon identification down to $p_{\rm T}\sim1.5$ GeV/$c$. The baseline option for the MID consists of a chamber with two layers of plastic scintillator bars equipped with wavelength-shifting fibers and read out with silicon photomultipliers (SiPMs). The layers are placed perpendicular to each other to form a grid-like structure. 

To test the developed muon-tagging algorithm, compare the background rejection with that obtained in simulations, and evaluate the technologies considered for the detector, a series of test-beam campaigns was performed at the CERN Proton Synchrotron using two different prototypes based on FNAL-NICADD scintillator bars equipped with Kuraray wavelength-shifting fibers and Hamamatsu SiPMs for readout.

\section{Results}

The first prototype consisted of a chamber with an active area of $21.4 \times 21.4$~cm$^2$, built from two perpendicular layers of five $25 \times 4 \times 1$~cm$^3$ scintillator paddles separated by 10~cm. The experimental setup included a scintillator array to obtain a point-like beam and MWPCs placed before and after an iron absorber of variable thickness to measure the profile of the 3~GeV/$c$ muon- and pion-enriched beams used in the test.

\begin{figure}[t]
    \centering
    \includegraphics[width=0.4\textwidth]{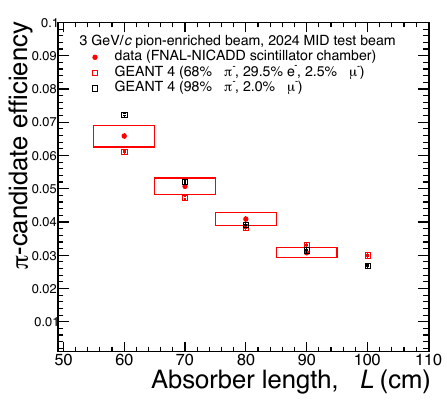}
    \includegraphics[width=0.4\textwidth]{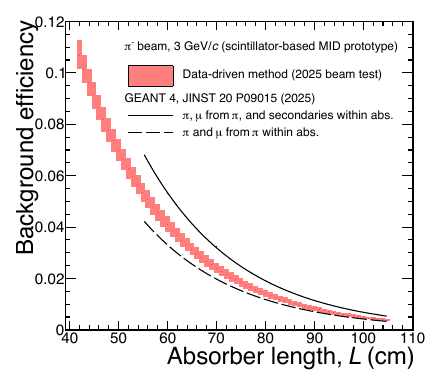}
    \caption{Left: Pion-candidate efficiency versus absorber length for the smaller prototype, in data and MC. Lines show statistical uncertainties; boxes show the $\sim5\%$ systematic uncertainty from the efficiency difference between the MID chamber and the MWPCs. Right: Background efficiency versus absorber length for the larger prototype, in data (in red) and for two beam compositions in MC (black lines). The red band shows the systematic uncertainty from the 2~cm hole in the absorber during this test beam. Figures are taken from Ref.~\cite{MunozMendez:2025ttk} and~\cite{AlfaroMolina:2026asj}, respectively. }
    \label{fig:MID_results}
\end{figure}

The data analysis relied on boosted decision trees (BDT) trained using GEANT4 simulations of the MID chamber, with the measured variables as input features. The pion-candidate efficiency as a function of the absorber length, obtained considering the trained BDT, is shown in the left panel of Fig.~\ref{fig:MID_results}, for both data and simulations. Two different beam compositions were considered in the simulations: one with only pion-decay muon contamination, and another with 29.5\% electrons and 2.5\% muons, which is closer to the reported beam composition \cite{vanDijk:2025ggb}. Both show good agreement with the data. For the baseline 70~cm absorber, the pion-candidate efficiency of the enriched beam is $5 \pm 0.5\%$, with the uncertainty assigned from the consistency between the results of the scintillator chamber and the MWPC array . However, in a pure-pion beam simulation, muons account for about 15\% of this efficiency and other secondaries for 34\%, yielding a pure-pion efficiency of $3.0 \pm 0.15\%$. The latter uncertainty is taken from the difference between the enriched and pure beams.

The second prototype consisted of 24 paddles of $100 \times 4 \times 1$~cm$^3$ per layer, a layer separation of 1~cm and a bar spacing of 0.1~cm, increasing both the active area and the geometrical acceptance. In this test, a high-pressure threshold Cherenkov counter was placed along the beamline to reduce the electron contamination. 

For this analysis, and additionally to the BDT, a template-fit method was applied to isolate the primary-candidate fraction from the secondaries, which further suppresses the background. As a consequence, the resulting data-driven background efficiency remains below the simulation which includes all background contributions, as shown in the right panel of Fig.~\ref{fig:MID_results}. With this prototype, the BDT-output threshold is set to yield a muon efficiency of 94\%, which corresponds to a background efficiency of $2.4 \pm 0.1\%$ for a 70~cm absorber, compatible with the result from the smaller prototype.

\section{Summary}

Two MID prototypes based on FNAL-NICADD scintillator bars, Kuraray fibers and Hamamatsu SiPMs were characterized. For the baseline 70~cm absorber, a muon efficiency of 94\% at a background efficiency of $2.4\pm0.1\%$ was measured, compatible with the simulations reported in~\cite{ALICE:2022wwr,Dainese:2925455} and sufficient for the physics program to be fulfilled. These results show that the proposed configuration of plastic scintillators equipped with SiPMs is a suitable, robust, and cost-effective solution for the ALICE 3 upgrade.\\

\noindent\textbf{Acknowledgments.} This work has been supported by DGAPA-UNAM PAPIIT No. IG100524

\end{document}